# Phishing Techniques in Mobile Devices

**Belal Amro**

College of Information Technology, Hebron University, Hebron, Palestine
Email: bilala@hebron.edu

## Abstract

The rapid evolution in mobile devices and communication technology has increased the number of mobile device users dramatically. The mobile device has replaced many other devices and is used to perform many tasks ranging from establishing a phone call to performing critical and sensitive tasks like money payments. Since the mobile device is accompanying a person most of his time, it is highly probably that it includes personal and sensitive data for that person. The increased use of mobile devices in daily life made mobile systems an excellent target for attacks. One of the most important attacks is phishing attack in which an attacker tries to get the credential of the victim and impersonate him. In this paper, analysis of different types of phishing attacks on mobile devices is provided. Mitigation techniques—anti-phishing techniques—are also analyzed. Assessment of each technique and a summary of its advantages and disadvantages is provided. At the end, important steps to guard against phishing attacks are provided. The aim of the work is to put phishing attacks on mobile systems in light, and to make people aware of these attacks and how to avoid them.

## 1. Introduction

During the last 10 years, mobile devices technologies have grown rapidly due to the daily increase in the number of users and facilities. According to [1], the number of mobile users has become 4.92 billion global users in 2017. Current mobile devices can be used for many private and financial applications as Facebook, mobile bank, etc. Android and iOS are the two dominant operating systems with 99.6% market share distributed as 81.7% for Android and 17.9% for iOS [2].

According to Symantec, phishing is defined as "an attempt to illegally gather personal and financial information by sending a message that appears to be from a well-known and trusted company". The phishing message has a fake link to a crafted webpage similar to the legitimate page, the user is asked to provide his credentials to log into the page which causes his credentials to be transferred to the phisher. To stay hidden and un-noticed, the fake webpage then redirects the user to the legitimate page.

According to [3], at least 255,065 unique phishing attacks occurred worldwide. The increase is over 10% from the attacks identified during 2015. The distribution of these attacks on industry is shown in Figure 1.

Due to the sensitivity of the data stored on mobile devices, these devices have become an excellent target for phisher to launch their attacks. The aim of the attacks is to gain access to credential that might benefit of using services a user is registered to. These services include dialing, SMS, payments, sensitive data leakage and connectivity. A phisher might impersonate the mobile user and use his mobile to perform these tasks without authorization from the user.

According to [4], more than 4000 ransomware attacks occurred daily during the year of 2016. PhishMeInc reported that ransomware and phishing attacks work together and that 97.2 of phishing emails in 2016 contain a form of ransomware [5]. Figure 2 shows the frequency of a ransomware attack on individuals in Q1 and Q3 of the year 2016. It is clearly shown that the frequency of the attack on individuals has doubled in Q3.

A study has been conducted by Dr. Zinaida Benenson at Friedrich-Alexander University (FAU) showed that 78% of people showed awareness of phishing attacks with unknown links and 45% of them clicked the link [6].

According to the numbers and figures showed here we conclude that phishing attacks on mobile devices are increasing dramatically. Besides, attacks have spread to cover wide areas of services as shown in Figure 1. And bearing into mind that mobile users are not aware of this huge threat—and if so they still follow malicious links—it is very important to spot the light on this type of phishing attacks and to make mobile users ware of the size of the threat and the steps they might follow to mitigate this attack.

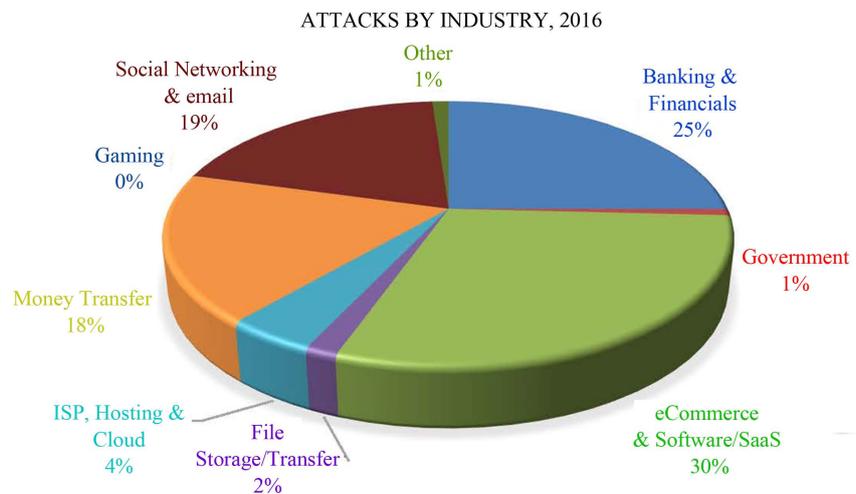

Figure 1. Distribution of phishing attacks over industry. Source: APWG report on global phishing trends and domain name use 2016.

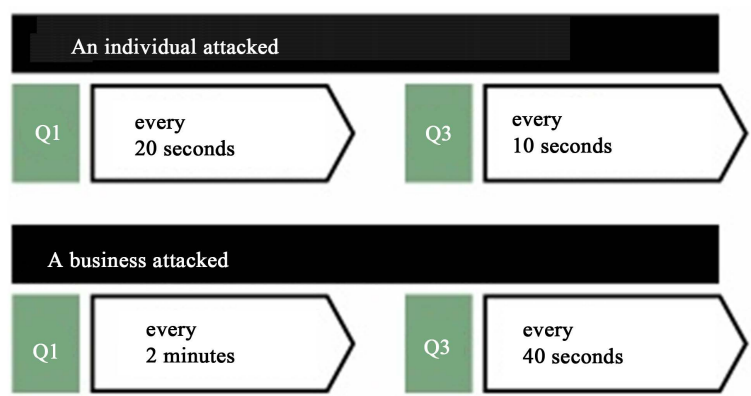

Figure 2. Frequency of ransomware attacks on individuals in 2016.

The rest of the paper is organized as follows, mobile phishing attacks are explained in Section 2. Section 3 details Mobile anti-phishing techniques. Steps required to avoid phishing attacks are provided in Section 4. Section 5 summarizes the work done in this paper.

## 2. Mobile Phishing Attacks

Before talking about mobile phishing attacks, we have first to introduce the attractive properties of mobile devices that made malware creators target these devices, then we will talk about types of phishing attacks. Finally we will talk about distribution methods used in mobile applications with some statistics.

### 2.1. Properties of Mobile Device Usage

Mobile devices facilitate phishing attacks due to their following properties:

The rapid increase of the number of mobile users worldwide as describes in [1]. This user shift to mobile devices has attracted phisher to shift their technique to mobile devices.

The limited screen size makes it difficult for mobile users to pinpoint legitimate webpage from phishing one. Besides, the small screen size also makes browsers to hide the complete URL of the requested page, and hence help the phisher deceiving the mobile user.

Due to the mobility nature of the mobile usage, users tend to respond to interactions with less concentration which might yield to approving a phishing process.

Because the mobile device is mostly near the mobile users, users tend to trust these devices. This in turn will leverage the possibility of being hacked and phished due to this trust.

Unfortunately, malware creators and phishers are aware of these attractive properties and hence have moved their efforts and techniques to mobile devices and their applications.

## 2.2. Methods of Mobile Phishing Attacks

The aim of phishing is to acquire credential that might be used to impersonate the person using his credentials. The basic idea to have a successful phishing attack is to deceive the user to provide his credentials. In mobile technology, there are different ways fishers used to launch their fishing attacks and deceive victims, these methods are listed below:

### 2.2.1. Financial Fraud

As the name implies, financial fraud aims to gather financial credentials of a victim, these credentials might then be used to impersonate the person and perform financial transactions on behalf of him. The report of MicroSave [7], provides details about financial mobile frauds. The report states that mobile financial fraud is becoming increasingly important with the extended use of financial mobile applications to perform electronic financial transactions.

CGAP [8] has conducted a research about financial fraud in different countries. The key idea was that it is not possible to completely defend against this fraud. However, following some steps by providers with hints we will provide might help mitigating this fraud.

One example of financial fraud might be a false update for the internet banking account. The user follows the link in the message which looks like legitimate and he will be directed to a page similar to the login page of his bank. To avoid such a scenario, service providers like banks usually perform two factor authentication techniques typically send a pin code as an SMS to the mobile of the user. If the entered PIN is not identical to the one sent, then login is prohibited.

### 2.2.2. Service Updates

In this method of attacks, fishers use registered services for users to collect their credentials and hence impersonate legitimate users and use the services instead. Different services are available over internet including Drop Box, Google Drive, Microsoft Account, etc. A user receives a message showing the necessity of service update, the user is required to supply his credential for the service to be updated. This in turn handles the credentials associated with this service to the fisher.

As in financial fraud, some service providers use two factor authentications. However, a fisher might bypass this by being online when launching the attack and will also receive the PIN code as well.

### 2.2.3. Promotional Offers

A fisher sends fake promotional offers automatically to a number of users, these offers are in the form of purchasing coupons, tickets, gifts, etc. the user is asked to create an account to get that offer. The fisher then uses the provided credentials to try to login into other systems hoping that the user has used the same credentials.

### 2.2.4. Spear Phishing

This type of attack is more targeted to a person or an organization. This type of attack requires social engineering to gather correct data and then deceive the target person being someone who knows. This in turn enables the fisher to extract information from that person pretending to be someone who knows.

An example of a real-life sphere attack reported by [9]. In 2010, a spear attack involved going after code on many machines using malware to access Google, Adobe, and other U.S. systems. The attack aimed at stealing intellectual property to an Executive Editor at https://www.darkreading.com/.

### 2.2.5. Whaling

Whaling is a special type of sphere phishing. The victim of the attack is of high profile person. A hacker spends large time gathering information about his target, when sufficient information is available, the attacker will use his information to launch the attack. The attacker might need some time to build trust with his victim before starting his attack. Once trust is good enough, the attacker can exploit this trust and perform his attack.

An example of whaling attack is reported by [10] where Leoni AG, Europe's biggest manufacturer of wires and electrical cables, lost $44.6 million as a result of a whaling attack that tricked finance staff into transferring money to the wrong bank account.

### 2.3. Spreading Methods

By spreading method, we mean the origin of the mobile traffic used for phishing. According to [8], iOS is more targeted to phishing attacks than Android. The percentage of phishing attacks of iOS is 63% while it is only 37% for android. The justification is that Apple users are more prestigious and hence are better phishing targets than others.

There are many distribution techniques used for phishing. The popularity of these techniques might be different in mobile application compared to other applications. Table 1 below summarizes the popularity of spreading techniques according to [11].

As seen far, mobile phishing techniques are becoming more dedicated and technically well organized. The targets of phishing are becoming more specific and selected well. Social engineering is used intensively to make the phishing more effective. In the next section, we will talk about mobile anti-phishing techniques and their advantages and disadvantages.

## 3. Mobile Anti-Phishing Techniques

It is worth mentioning here that phishing techniques in mobile applications can be done via a web browser or using a login page for a particular mobile application. Anti-phishing technique must be able to detect phishing attacks via browsers or applications. Different web anti-phishing techniques has been proposed, these techniques are content based techniques, black lists, and white lists. In content based techniques the content of the website is used to determine whether it is a phishing site or not [12]. While black lists and white lists compares the requested URL against either a black list of phishing URLs, or a white list of non-phishing ones [13] [14][15].

Table 1. Phishing spreading methods.

| Spreading technique | Percentage % |
|---|---|
| Games | 25.6 |
| Email | 18.9 |
| Sports | 13.3 |
| News and weather | 13.1 |
| Productivity | 9.4 |
| Social media | 8.1 |
| Messaging | 6.4 |
| Travel | 6.1 |
| Ecommerce | 5.8 |
| Music | 5.6 |
| Dating | 5.0 |
| Others | 3.9 |

In [16], the authors proposed a method built upon having the users registered to a website, the later generates a unique code for each user. The user is asked to input some digits of his unique code and the website has to respond with the complete code. Correct code means that the website is genuine. This method requires users to sign up and remember their code for different websites, furthermore, once a code is compromised, the phishing will easily be launched.

A framework for detection of phishing is provided in [17]. The proposed framework extracts features of a website and compares them to the original features of the genuine site. However, most phishing pages are very similar to the genuine ones and hence this might produce false positives. The framework also requires high computation to extract text, image, and color features of a website.

A user visual similarity is proposed in [18]. The method uses screenshot extraction to calculate the visual similarities. The authors proposed a metric called deception rate for calculating similarity. The method is dedicated for mobile apps login screens and does not work for web logins.

A naïve Bayesian method is proposed in [19]. The method is based on building a learning model based on gathering data about permissions and key logs. After learning, the model is used to check applications accordingly. However, the system requires upgrades regarding memory and control scheme, after that it might be evaluated for practically being used.

In [20], a light weight technique is used. It uses some URL features to determine legitimacy of a webpage based on frequency analysis of extracted features of phishing URLs. The method is not able to detect for phishing sites of new behaviors with respect to phishing URL behavior depository.

Visual cryptography is used for phishing detection [21]. The proposed framework is an interactive method that uses captcha validation scheme based on visual cryptography. Both user and server own part of a captcha and both work together to reconstruct the complete captcha image. The time spent for captcha reconstruction is recorded and used in the process of phishing detection. The framework is vulnerable to man-in-the-middle attack. Besides, it requires the user to share the secret with the server.

In [22], an MP-shield method is proposed. The method relies on searching Google for blacklisted URL, besides it extracts features of URL and passes them to the proper classifier model. The method might generate false positives and relies on Google for internet search and not on a dedicated trusted blacklist.

## 4. Avoiding Phishing

To avoid phishing a user must be aware of the URL, the URL can give an indication of being a target of a phishing attack. However, as we mentioned before, the small screen of a mobile device make it difficult for users to check for URLs. Here we provide some hints that might be useful for mobile users to avoid phishing attacks.

1) The most important issue is to install and use genuine applications provided by trusted vendors. This will make a user trust his applications of not being part of any phishing attack. According to [23], PhishLabs company reported that 11 malicious applications claiming to be genuine mobile payment apps were on the official Google Play application store.

2) A user might also use an anti-phishing tool from trusted vendors, this will enable him to detect phishing attacks and avoid them.

3) Never reply to any suspicious emails or SMSs. If you are suspicious of why this is delivered to you then you might be at high risk of being a target of phishing attack. Do not reply to such messages at all.

4) A good practice to avoid phishing is to use bookmarks for your frequently visited websites. This will make it difficult to land unwanted webpages that might be phishing sites.

5) A user must follow up security notes and train himself on secure mobile usage.

6) It is important to highly secure a mobile device by a password and other access control methods. This will protect identity theft once the mobile device is lost.

7) Use Anti-Theft security services like Remote Lock, Remote Wipe, and Locate to protect your device and the data it contains.

8) Use safe browsing techniques to avoid visiting malicious websites.

## 5. Summary

In this paper, we provided a summary of anti-phishing techniques used for mobile device. These techniques relied on some features of mobile devices that made phishing more likely to happen. Some of these features were small screen size and the feel of trust by most users in mobile devices. We have explained anti-phishing techniques and methods that were proposed for mobile devices. These techniques have their pros and cons. Our investigation showed that all these techniques have some shortcomings that reduce their efficiency in detecting phishing attacks. Hence, the most burden as we believe falls onto mobile users to follow some steps that might help them avoid phishing. These steps and best practices were also provided in the last section of our paper. The aim of our paper as mentioned in the abstract is to put phishing attacks on mobile systems in light, and to make people aware of these attacks and how to avoid them.

## References

[1] https://wearesocial.com/special-reports/digital-in-2017-global-overview

[2] http://www.gartner.com/newsroom/id/3609817

[3] Aijaz Ahmad, S., *et al.* (2013) Smartphone: Android vs IOS. *The SIJ Transactions on Computer Science Engineering* & *Its Applications* (*CSEA*), **1**, 141-148.

[4] US Government (2016) How to Protect Your Network from Ransomware. Technical Guidance Interagency Document. US Government, Washington, DC.

[5] PhishMe, Inc. (2016) Malware Review Q3. PhishMe, Inc., Leesburg, VA.

[6] https://www.fau.eu/2016/08/25/news/research/one-in-two-users-click-on-links-from-unknown-senders/

[7] Luminzu Mudiri, J. (2012) Fraud in Mobile Financial Services. MicroSave, Lucknow.

[8] Buku, M.W. and Mazer, R. (2015) Fraud in Mobile Financial Services: Protecting Consumers, Providers, and the System. CGAP, Washington, DC.

[9] http://resources.infosecinstitute.com/spear-phishing-real-life-examples/#gref

[10] https://www.scmagazineuk.com/leoni-ag-suffers-34-million-whaling-attack/article/530694/

[11] WANDERA (2017) Mobile Data Report: Focus on Phishing.

[12] Yoon, J.W., *et al.* (2010) Hybrid Spam Filtering for Mobile Communication. *Computers and Security*, **29**, 446-459. https://doi.org/10.1016/j.cose.2009.11.003

[13] Memon, I.K. and Khan, M.K. (2013) Anti Phishing for Mid-Range Mobile Phones. *International Journal of Computer and Communication Engineering*, **2**, 115-119.

[14] Singh, D., *et al.* (2011) Telephony Fraud Prevention. US Patent.

[15] Mahmoud, T.M. and Mahfouz, A.M. (2012) SMS Spam Filtering Technique Based on Artificial Immune System. *IJCSI International Journal of Computer Science Issues*, **9**, 589-597.

[16] Mishra, M., *et al.* (2012) A Preventive Anti-Phishing Technique using Code Word. *International Journal of Computer Science and Information Technologies*, **3**, 4248-4250.

[17] Archana, M., *et al.* (2011) Architecture for the Detection of Phishing in Mobile In-

ternet. *International Journal of Computer Science and Information Technologies*, **2**, 1297-1299.

[18] Malisa, L., *et al.* (2015) Technical Report: Detecting Mobile Application Spoofing Attacks by Leveraging User Visual Similarity Perception.

[19] Kumar, N. and Chaudhary, P. (2017) Mobile Phishing Detection using Naive Bayesian Algorithm. *International Journal of Computer Science and Network Security*, **17**, 142-147.

[20] Orunsolu, A.A. (2017) A Lightweight Anti-Phishing Technique for Mobile Phone. *Acta Informatica Pragensia*, **6**, 114-123.

[21] Yenurkar, B. and Zade, S. (2014) An Anti-Phishing Framework with New Validation Scheme Using Visual Cryptography. *International Journal of Computer Science and Mobile Computing*, **3**, 739-744.

[22] Bottazzi, G. (2015) MP-Shield: A Framework for Phishing Detection in Mobile Devices. *IEEE International Conference on Computer and Information Technology*; *Ubiquitous Computing and Communications*; *Dependable, Autonomic and Secure Computing*; *Pervasive Intelligence and Computing*, Liverpool, 26-28 October 2015, 1977-1983. https://doi.org/10.1109/CIT/IUCC/DASC/PICOM.2015.293

[23] http://onlinesecurity.trendmicro.com.au/blog/2016/06/22/phishlabs-warns-of-malware-posing-as-legit-apps-on-google-play/